# Adapting Altman's Bankruptcy Prediction Model to the Compositional Data Methodology

**Fatemeh Keivani**
University of Girona
Girona, Spain

**Germà Coenders**
University of Girona
Girona, Spain

**Geòrgia Escaramís**
University of Girona
Girona, Spain
CEEISCAT,
Department of Health
Government of Catalonia
Barcelona, Spain

**Abstract**

Using standard financial ratios as variables in statistical analyses has been related to several serious problems, such as extreme outliers, asymmetry, non-normality, and non-linearity. The compositional-data methodology has been successfully applied to solve these problems and has always yielded substantially different results when compared to standard financial ratios. An under-researched area is the use of financial log-ratios computed with the compositional-data methodology to predict bankruptcy or the related terms of business default, insolvency or failure. Another under-researched area is the use of machine learning methods in combination with compositional log-ratios. The present article adapts the classical Altman bankruptcy prediction model and some of its extensions to the compositional methodology with pairwise log-ratios and three common statistical and machine learning tools: logistic regression models, k-nearest neighbours, and random forests, and compares the results with standard financial ratios. Data from the sector in the Spanish economy with the largest number of bankrupt firms according to the first two digits of the NACE code (46XX "wholesale trade, except of motor vehicles and motorcycles") were obtained from the Iberian Balance sheet Analysis System. The sample size (31,131 firms, of which 97 were bankrupt) was divided into a training and a validation dataset. The training dataset was downsampled to one healthy firm to each bankrupt firm. No outliers were removed. Focusing on predictive performance, the results show that compositional methods are better than standard ratios in terms of sensitivity (recall), with mixed results regarding specificity, compositional random forests and compositional logistic regression behaving the best.

## 1. Introduction

Using standard financial ratios as variables in statistical models has been related to severe violations of the models' assumptions, such as *extreme outliers, asymmetry*, connected to it *severe non-normality* of the distributions, *non-linearity* of the relationships, and even dependence of the results on the arbitrary decision regarding which accounting figure appears in the numerator and which in the denominator of the ratio (Carreras-Simó and Coenders 2021; Coenders and Arimany Serrat 2025b; Durana, Kovalova, Blazek, and Bicanovska 2025; Iotti, Ferri, and Bonazzi 2024; Iotti, Ferri, and Calugi 2026; Linares-Mustarós, Farreras-Noguer, Arimany-Serrat, and Coenders 2022; Magrini 2025). The results of many statistical and *machine learning* analyses are invalid when all or some of these problems occur and the results and conclusions of said analyses can be affected to a great extent. The well-known *Compositional Data* (CoDa) methodology using log-ratio transformations (Aitchison 1982, 1986; Coenders, Egozcue, Fačevicová, Navarro-López, Palarea-Albaladejo, Pawlowsky-Glahn, and Tolosana-Delgado 2023a; Greenacre, Grunsky, Bacon-Shone, Erb, and Quinn 2023; Pawlowsky-Glahn, Egozcue, and Tolosana-Delgado 2015) has been successfully applied in connection with financial ratios over the last 10 years, to solve these problems (Coenders 2025; Coenders and Arimany Serrat 2025a,b; Kuryłek and Kowalewski 2026) This methodology was developed to deal with fixed-sum data in chemical and geological analysis, but the fixed sum is by no means essential. CoDa are contemporarily defined as arrays of strictly positive numbers for which ratios between them are considered to be relevant (Egozcue and Pawlowsky-Glahn 2019), which perfectly fits the notion of financial statement analysis.

Far from being a statistical refinement, the CoDa methodology leads to very substantial differences in the analysis results whenever it has been compared with standard financial ratios (Arimany-Serrat, Farreras-Noguer, and Coenders 2022; Carreras-Simó and Coenders 2021; Coenders and Arimany Serrat 2025a,b; Coenders, Sgorla, Arimany-Serrat, Linares-Mustarós, and Farreras-Noguer 2023b; Creixans-Tenas, Coenders, and Arimany-Serrat 2019; Dao, Coenders, Lai, Dam, and Trinh 2026; Escaramís and Arbussà 2025; Hernandez Romero and Coenders 2025; Jofre-Campuzano and Coenders 2022; Linares-Mustarós, Coenders, and Vives-Mestres 2018; Linares-Mustarós *et al.* 2022; Magrini 2025; Molas-Colomer, Linares-Mustarós, Farreras-Noguer, and Ferrer-Comalat 2024).

An under-researched area is that of the use of financial-statement ratios to predict outcome variables (Coenders and Arimany Serrat 2025a; Molas-Colomer, Ferrer-Comalat, Carreras-Pijuan, and Linares-Mustarós 2026; Taunk and Patidar 2026) such as *bankruptcy* or the related terms of *business default*, *insolvency* or *failure*. The only known instance of application of the CoDa methodology here is that by Magrini (2025) using *logistic regression models*, *LASSO regularization*, and *random forests*. Magrini's work is also the first to combine a financial application of the CoDa methodology with machine learning. In bankruptcy prediction studies, predictive accuracy becomes a key parameter, besides fulfilment or violation of the assumptions. Magrini did not find substantial differences in accuracy between standard and compositional methods. In this article we extend on Magrini's work, as explained below.

Since the seminal work by Altman (1968), the problem of using financial ratios to predict bankruptcy has never ceased to attract interest, and many recent review and bibliometric studies are availalable to which the reader can resort (Gwalani, Teltumbde, and Gwalani 2025; Hanun and Ferdiani 2025; Insani, Pontoh, and Said 2024; Khedr, Jemmali, Imtiyaz, and Ramic-Brkic 2026; Nugroho and Dewayanto 2025; Pinto and Pereira 2026; Voicila Voicila and Urquia Grande 2026). The original Altman's model and some popular variants continue to be used and adapted to new machine learning tools, a.k.a.*statistical learning*, *data mining*, and *artificial intelligence* tools (Martin-Melero, Gomez-Martinez, Medrano-Garcia, and Hernández-Perlines 2025). Including and beyond Altman's approach, the literature on bankruptcy prediction with machine learning methods is growing apace, 39 relevant scholarly articles being identified when searching "machine learning" and "bankruptcy prediction" in the SCOPUS database only between January and September 2026.

Often, eliminating a vast amount of outliers (Balzano and Magrini 2025; Magrini 2025; Pawełek and Sadko 2026), even in the range of one third of the original sample (Martin-Melero *et al.* 2025), is the price to be paid for using standard financial ratios, as some machine learning methods are relatively immune to outliers, but not all of them.

Machine learning tools can be adapted to the CoDa methodology (Tolosana-Delgado, Talebi, Khodadadzadeh, and van den Boogaart 2019; Tran 2025). The present article compares Altman's classical bankruptcy prediction model and some of its extensions with the CoDa methodology and three common tools, a statistical model (logistic regression), a robust machine learning tool (random forests), and a non-robust machine learning tool (*k-nearest neighbours*). Rather than focusing on the violation of the assumptions, we focus on comparing predictive performance. Our approach complements that of Magrini (2025), firstly, by using the full sample without trimming outliers for all the methods compared, which is possible in the CoDa methodology, as it tends to produce few or no outliers (Linares-Mustarós *et al.* 2018). Random forests are relative immune to outliers (Hia, Kuswanto, and Prastyo 2023), but logistic regression, being a parametric statistical model, and k-nearest neighbours, being a distance-based method, are not. Secondly, by downsampling healthy firms, which is known to improve predictive performance when there are few bankrupt firms (Antar and Tayachi 2025; Bonelli 2026; Gabrielli, Melioli, and Bertini 2026; Gnip, Kanász, Zoričak, and Drotár 2025; Kanojia and Arora 2025; Liu, Li, and Yu 2025; Martin-Melero *et al.* 2025; Tanaka, Higashide, Kinkyo, and Hamori 2025a,b; Veganzones, Séverin, and Ben Jabeur 2026). Thirdly, by using an alternative log-ratio transformation termed *pairwise log-ratios*, henceforth plr (Greenacre 2019), which has been shown to perform better with random forests (Tolosana-Delgado *et al.* 2019). Fourthly, by focusing on Altman's model and its derivatives (Martin-Melero *et al.* 2025). Fiftly, by using data from a homogeneous sector rather than a country's whole economy in order to reduce firm heterogeneity (Balzano and Magrini 2025).

# 2. Materials and methods

## 2.1. Standard and CoDa representation of Altman's model for bankruptcy prediction

According to Altman (1968), the following five standard financial ratios are good predictors of bankruptcy:

$$\begin{array}{ll}
\text{working capital ratio :} & (CA-CL)/(NCA+CA), \\
\text{retained earnings over assets :} & RE/(NCA+CA), \\
\text{return on assets :} & (OR-OE)/(NCA+CA), \\
\text{net worth over liabilities :} & (NCA+CA-NCL-CL)/(NCL+CL), \\
\text{turnover :} & OR/(NCA+CA),
\end{array} \tag{1}$$

where NCA stands for non-current assets, CA for current assets, RE for retained earnings, NCL for non-current liabilities, CL for current liabilities, OR for operating revenue and OE for operating expenses. In the original Altman's model, the company market value is used instead of the book value of net worth (NCA+CA–NCL–CL). The book value is used in this article as commonly done, being available for all companies, be they traded in the stock exchange or not.

Altman's model has been reformulated many times and continues to attract academic interest (Martin-Melero *et al.* 2025). The most popular such reformulations are arguably those by Amat i Salas, Martínez, and Roure (2008); Grover and Lavin (2001); Springate (1978); and Zmijewski (1983). These formulations add or substitute five more standard financial ratios:

$$\begin{array}{ll}
\text{profit over current liabilities :} & (OR-OE)/CL, \\
\text{current ratio :} & CA/CL, \\
\text{inverted leverage ratio :} & (NCA+CA-NCL-CL)/(NCA+CA), \\
\text{return on equity :} & (OR-OE)/(NCA+CA-NCL-CL), \\
\text{indebtedness :} & (NCL+CL)/(NCA+CA).
\end{array} \tag{2}$$

The starting point of the CoDa methodology is not the ratios, but a composition of the $D$ positive accounting figures needed to compute them, which are called *parts* or *components*. Positiveness is essential. For instance, one cannot use operating profit (OR–OE) which may be negative, but the always positive operating revenue (OR) and operating expenses (OE) separately (Creixans-Tenas *et al.* 2019). In our case, the $D = 7$ parts in the composition are NCA, CA, RE, NCL, CL, OR and OE.

The $D$ parts are then subject to log-ratio transformations that contain all the relative information among them (Coenders and Arimany Serrat 2025b; Linares-Mustarós *et al.* 2018).

The simplest approach to the CoDa methodology in financial analysis is transforming the accounting figures by means of plr (Creixans-Tenas *et al.* 2019; Coenders and Arimany Serrat 2025b; Farreras-Noguer, Carreras-Pijuan, Linares-Mustarós, and Arimany-Serrat 2025; Greenacre 2019; Mulet-Forteza, Ferrer-Rosell, Cunill, and Linares-Mustarós 2024). This involves defining $D-1$ log-ratios between only two accounting figures, so that each accounting figure can be connected to any other (exhaustiveness) in only one way (non-redundancy) through the log-ratios. The $D-1=6$ plr below follow both rules and can be interpreted according to common financial concepts:

$$\begin{array}{ll} \text{asset tangibility :} & \log\left(NCA/CA\right), \\ \text{current} - \text{asset turnover :} & \log\left(OR/CA\right), \\ \text{margin :} & \log\left(OR/OE\right), \\ \text{current ratio :} & \log\left(CA/CL\right), \\ \text{debt maturity :} & \log\left(NCL/CL\right), \\ \text{retained earnings over non} - \text{current liabilities :} & \log\left(RE/NCL\right). \end{array} \tag{3}$$

As recommended by Greenacre (2019), if we draw a graph in which the accounting figures are the vertices (nodes) and the plr are the connections (edges), it is easy to see that any accounting figure can be joined to any other through the plr in only one way, so that the graph is connected and acyclic. For instance, NCA and OR are connected to each other only through the $\log\left(NCA/CA\right)$ and $\log\left(OR/CA\right)$ plr. Edges can be drawn as arrows without affecting the graph properties, the arrows pointing at the numerator of the plr for clarification purposes (Figure 1).

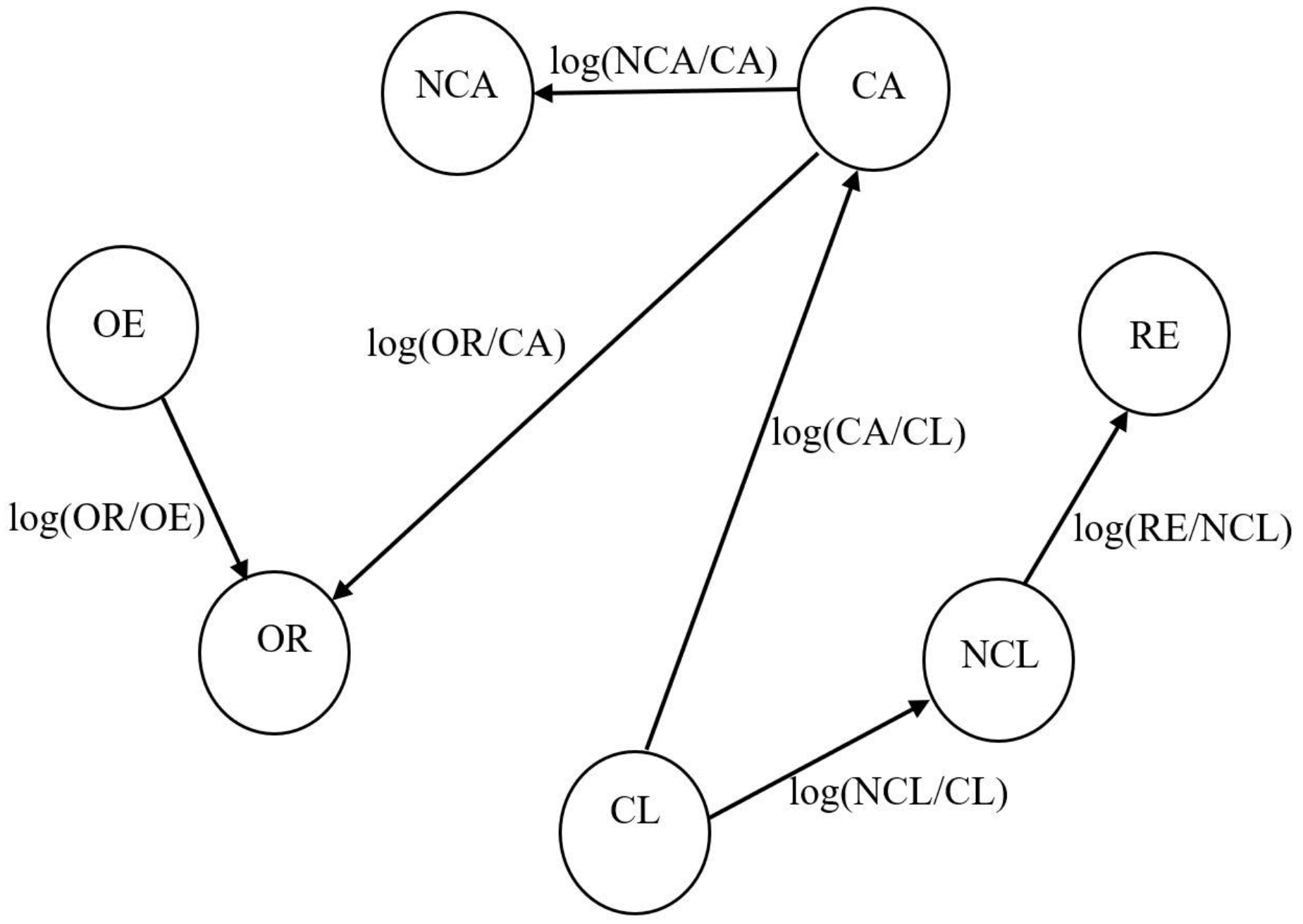

Figure 1: Connected acyclic graph among the $D = 7$ accounting figures

$D-1$ plr can routinely be used as a set of predictors in statistical models, including logistic regression models (Coenders and Greenacre 2023; Coenders and Pawlowsky-Glahn 2020). Furthermore, the predictions are invariant to the particular choice of $D-1$ plr, as long as each accounting figure can be connected to any other in only one way (Coenders and Pawlowsky-Glahn 2020). More than $D-1$ plr would lead to perfect collinearity.

The distances among companies do depend on the plr choice. Rather than selecting $D-1$ plr, using all the possible $D\,(D-1)\,/2$ plr is a better option for methods based on distances, such as the k-nearest-neighbour method, because Euclidean distances computed on all the possible $D\,(D-1)\,/2$ plr are equivalent to the standard distance measure for CoDa, termed *Aitchison's distance* (Aitchison 1983; Aitchison, Barceló-Vidal, Martín-Fernández, and Pawlowsky-Glahn 2000). Methods which have the ability to select the best predictors from a large available set, such as random forests, will also benefit from using all the possible $D\,(D-1)\,/2$ plr (Tolosana-Delgado *et al.* 2019). Besides, with random forests and k-nearest neighbours, the predictions depend on the plr choice when using only of $D-1$ of them, which makes results arbitrary and non-replicable.

Which accounting figure goes to the numerator or the denominator of the plr does not change the results. In our case, the set of $D\,(D-1)\,/2 = 21$ possible plr is:

$$\begin{array}{lll}
\log\left(NCA/CA\right), & \log\left(CA/NCL\right), & \log\left(RE/OE\right), \\
\log\left(NCA/RE\right), & \log\left(CA/CL\right), & \log\left(NCL/CL\right), \\
\log\left(NCA/NCL\right), & \log\left(OR/CA\right), & \log\left(NCL/OR\right), \\
\log\left(NCA/CL\right), & \log\left(CA/OE\right), & \log\left(NCL/OE\right), \\
\log\left(NCA/OR\right), & \log\left(RE/NCL\right), & \log\left(CL/OR\right), \\
\log\left(NCA/OE\right), & \log\left(RE/CL\right), & \log\left(CL/OE\right), \\
\log\left(CA/RE\right), & \log\left(RE/OR\right), & \log\left(OR/OE\right).
\end{array} \tag{4}$$

While the log-ratios in Equation (3) are meant for financial interpretation, those in Equation (4) are only constructed for equivalence with Aitchison's distances (k-nearest neighbours) and optimal predictive performance (random forests).

Thus, the 10 predictors in Equations (1) and (2) were used for the standard-ratio approach under all three prediction methods. The 6 predictors in Equation (3) were used for logistic regression under the compositional approach, and the 21 predictors in Equation (4) were used for random forests and k-nearest neighbours under the compositional approach. As under the original Altman's model and cited extensions, no control variables were included.

## 2.2. Statistical analysis

The logistic regression model was fitted with the `glm` function in the basis R software (R Core Team 2025). The k-nearest-neighbour classification was carried out with the `knn` function, which is part of the `class` R library (Venables and Ripley 2010) with $k = 5$ and without standardizing the plr, as required by Aitchison's distances. The random-forest classification used the `randomForest` function in the homonymous R library (Breiman 2001), by generating 100 trees and randomly selecting 4 predictors in the standard-ratio approach and 7 in the compositional approach, which is about one third of the 10 standard and 21 compositional ratios, respectively.

The sample was randomly divided into a *training subset* (70 % of the sample) and a *validation subset* to assess predictive quality (30 % of the sample). The validation subset was kept entirely and was not used in model estimation. To assess the stability of predictive performance, repeated stratified 5-fold cross-validation was performed on the training subset. The procedure was repeated 10 times using different fold partitions, resulting in 50 resamples. Within each resample, four folds were used for model estimation and the remaining fold for validation. To address class imbalance, downsampling was applied exclusively to the four training folds by randomly selecting one healthy firm for each bankrupt firm. The validation fold was not downsampled and therefore retained the original class distribution. The same cross-validation partitions were used for all prediction methods and ratio representations.

Five common precision measures are:

- *predictive accuracy* (overall percentage of correct predictions),
- *sensitivity, a.k.a. recall* (percentage of true bankrupt companies predicted as so),
- *specificity* (percentage of true healthy companies predicted as so),
- *balanced accuracy* (arithmetic average of sensitivity and specificity),
- *area under the receiver operating characteristic (ROC) curve.* We computed predictive accuracy, sensitivity, specificity and balanced accuracy by classifying a company as bankrupt above a 0.5 bankruptcy probability cut-off. The ROC curve plots sensitivity against specificity for all possible choices of the probability cut-off. The area under the curve (AUC) is 1 for perfect prediction and 0.5 for random guess. An optimal cut-off could also be computed from the ROC curve, but in our example it was close to 0.5 for all methods.

To allow a direct comparison of variable importance across the three classification methods, permutation variable importance was used. The values of each predictor were randomly permuted in the held-out validation sample while keeping all other predictors unchanged. Variable importance was measured as the decrease in the AUC relative to the original, unpermuted validation data. The permutation procedure was repeated 100 times for each predictor, and the mean decrease in AUC was used as the variable importance measure.

## 2.3. Data collection and pre-processing

We defined a company as being bankrupt if under *creditors' meeting* ("*concurso de acreedores*" in Spain). To predict if a company was bankrupt in the last available year, the standard and compositional ratio values of the year immediately before were used. The forecasting horizon and the number of lags for the predictor values are thus both 1.

Data gathering and pre-processing were as follows:

1. In order to balance data homogeneity (and hence comparability, Balzano and Magrini (2025)) and reasonable number of bankrupt firms, data from the sector in the Spanish economy with the largest number of firms under creditors' meetings in 2024 according to the first two digits of the NACE code (46XX "Wholesale trade, except of motor vehicles and motorcycles") were selected. Filters included having 5 or more employees and data available until at least 2023. The data were obtained on 6/10/2025 from the Iberian Balance sheet Analysis System (SABI, accessible at https://sabi.bvdinfo.com/) database, developed by INFORMA D&B in collaboration with Bureau Van Dijk.
2. The sample size was 31,131 firms, of which 97 were bankrupt. The sample breakdown was 2,796 firms for the NACE codes 461X "Wholesale on a fee or contract basis", 1,425 firms for 462X "Wholesale of agricultural raw materials and live animals", 8,775 firms for 463X "Wholesale of food, beverages and tobacco", 5,130 firms for 464X "Wholesale of household goods", 1,263 firms for 465X "Wholesale of information and communication equipment", 3,511 firms for 466X "Wholesale of other machinery, equipment and supplies", 6,870 firms for 467X "Other specialised wholesale", 1,361 and firms for 469X "Non-specialised wholesale trade". Some key descriptive statistics (first quartile, median, and third quartile) were respectively 6, 10, and 20 for the number of employees; EUR 825,987, EUR 1,851,218, and EUR 4,433,838 for total assets; and EUR 1,351,922, EUR 2,881,756, and EUR 6,887,335 for operating revenue.

3. No outliers were removed, only inactive firms having either zero total assets (NCA+CA), zero OR, and zero OE during the last available year or the year before the last.

4. The remaining zeros were imputed with the log-ratio EM method (Palarea-Albaladejo and Martín-Fernández 2008) available in the freeware CoDaPack (Comas-Cufí and Thió-Henestrosa 2011) downloadable at https://ima.udg.edu/codapack/. The percentages of zeros were 1.38 % for NCA, 0.01 % for CA, 5.54 % for RE, 22.58 % for NCL, and 0.05 % for CL. These percentages are deemed suitable for robust replacement with this method (Martín-Fernández, Hron, Templ, Filzmoser, and Palarea-Albaladejo 2012) even if it does not account for uncertainty associated with the imputation procedure. Alternative approaches can be found in Faugeras (2026); Palarea-Albaladejo and Martín-Fernández (2015).

5. The sample was randomly divided into a training set (70 % of cases) and a validation set (30 % of cases). The training set had 21,791 firms, of which 69 were bankrupt. The validation set had 9,340 firms, of which 28 were bankrupt. The downsampled training set thus contained 69 healthy and 69 bankrupt firms, and the validation set 9,312 healthy and 28 bankrupt firms. The same sets were used for all three methods both for the standard and compositional approaches.

Table 1 shows the skewness and kurtosis statistics of the standard ratios in Equations (1) and (2), and the plr in Equation (4) for the full sample. Skewness serves as a measure of asymmetry and does not, per se, invalidate any of the methods considered in this article because none of them assumes normality. Kurtosis is a measure of heavy distribution tails, in other words, outliers. All standard ratios but one have extreme kurtosis, making them on the paper unfit for non-robust statistical models and machine learning methods (in our case logistic regression and k-nearest neighbours) except after pruning outliers. Pruning outliers following the standard criterion (Martin-Melero *et al.* 2025) of values above the third quartile plus 1.5×(interquartile range) or below the first quartile minus 1.5×(interquartile range) would result in dropping 9,544 cases from the full sample, which is 30.7 % of the companies, thus seriously impairing the sample representativeness.

In the case of plr, only the margin ratio $\log(OR/OE)$ has high kurtosis, but to a much lesser extent than the standard ratios. The rest of the plr behave well. In contrast with Magrini (2025), we analyse the full sample without pruning outliers both for the standard and compositional approaches.

## 3. Results

According to the logistic regression model with standard ratios, the relevant predictors (5 % significance) are $(NCA + CA - NCL - CL)/(NCL + CL)$ (estimated coefficient $-5.48$), $(OR - OE)/CL$, (estimated coefficient $-5.26$), and $CA/CL$ (estimated coefficient $+2.61$) but, due to outliers, some predicted probabilities are numerically equal to 0 and 1, which affects the reliability of precisely the standard errors and thus the significance measures. Furthermore, redundancy among standard ratios may prevent interpretation, as keeping the remaining ratios constant may be impossible. For instance, $(NCA + CA - NCL - CL)/(NCL + CL)$ cannot be changed while keeping constant $(NCA + CA - NCL - CL)/(NCA + CA)$.

With compositional ratios, no numerical estimation problems are encountered, and the statistically significant predictors are $\log(OR/CA)$ (estimated coefficient $-0.82$), and $\log(RE/NCL)$ (estimated coefficient $-0.30$). $D-1$ plr constructed from a connected acyclic graph are non-redundant by construction and their effects can be interpreted at constant values of the remaining plr. Since the dependent variable in a logistic regression model is the log-odds of a firm being bankrupt, the effect of a log-transformed predictor is readily interpreted as an elasticity. Thus, increasing current–asset turnover $(OR/CA)$ by 1 % while keeping the remaining ratios in Equation (3) constant leads to a 0.82 % decrease in the odds of a firm being bankrupt. Increasing retained earnings over non–current liabilities ($RE/NCL$) by 1 % while keeping the remaining ratios in Equation (3) constant leads to a 0.30 % decrease in the odds of a firm being bankrupt.

Table 1: Skewness and kurtosis of standard ratios and plr in the full sample

| | Skewness | Kurtosis |
|---|---|---|
| (CA–CL)/(NCA+CA) | -30.3 | 2220.3 |
| RE/(NCA+CA) | 1.3 | 4.9 |
| (OR–OE)/(NCA+CA) | -23.9 | 1843.6 |
| (NCA+CA–NCL–CL)/(NCL+CL) | 129.9 | 19809.4 |
| OR/(NCA+CA) | 63.9 | 6211.5 |
| (OR–OE)/CL | -167.6 | 29201.4 |
| CA/CL | 167.5 | 28935.9 |
| (NCA+CA–NCL–CL)/(NCA+CA) | -44.7 | 3617.9 |
| (OR–OE)/(NCA+CA–NCL–CL) | -161.2 | 27301.5 |
| (NCL+CL)/(NCA+CA) | 44.7 | 3617.9 |
| log(NCA/CA) | -0.9 | 2.8 |
| log(NCA/RE) | 0.8 | 2.8 |
| log(NCA/NCL) | 0.7 | 0.2 |
| log(NCA/CL) | -0.9 | 2.9 |
| log(NCA/OR) | -0.9 | 3.0 |
| log(NCA/OE) | -0.9 | 3.1 |
| log(CA/RE) | 1.8 | 3.4 |
| log(CA/NCL) | 0.8 | -0.4 |
| log(CA/CL) | 0.9 | 6.3 |
| log(OR/CA) | -0.2 | 5.7 |
| log(CA/OE) | 0.1 | 4.7 |
| log(RE/NCL) | 0.3 | 0.1 |
| log(RE/CL) | -1.2 | 2.0 |
| log(RE/OR) | -1.4 | 2.5 |
| log(RE/OE) | -1.4 | 2.5 |
| log(NCL/CL) | -0.8 | -0.3 |
| log(NCL/OR) | -0.8 | -0.3 |
| log(NCL/OE) | -0.8 | -0.3 |
| log(CL/OR) | -0.2 | 5.2 |
| log(CL/OE) | -0.3 | 6.1 |
| log(OR/OE) | -3.8 | 277.2 |

A common variable importance metric for all three methods can be computed by permutation as described in Section 2.2 (Figure 2). For logistic regression, the statistically significant variables are consistently among the five best. Even if the $D\,(D-1)\,/2$ log-ratios in Equation (4) are not always interpretable, sometimes ad-hoc interpretations can be provided. In this particular case, the best variables for k-nearest-neighbours and random-forests are $\log\,(NCL/OR)$, $\log\,(RE/NCL)$, and $\log\,(NCA/NCL)$, and are all connected to non-current liabilities and thus to long-term indebtedness. The best standard ratio $(NCA + CA - NCL - CL)/(NCL + CL)$ for logistic regression and k-nearest-neighbours is related to the same concept.

Table 2 contains the precision measures for all approaches. Compositional methods always have higher sensitivity than their standard counterparts. Random forests beat them also in specificity. Compositional and standard logistic regression have about the same specificity. Compositional k-nearest neighbours perform worst in terms of specificity. The overall assessment by the balanced accuracy is that compositional random forests and compositional logistic regression perform best. The intervals in Table 2 show that sensitivity and balanced accuracy have a large uncertainty due to the small number of bankrupt firms.

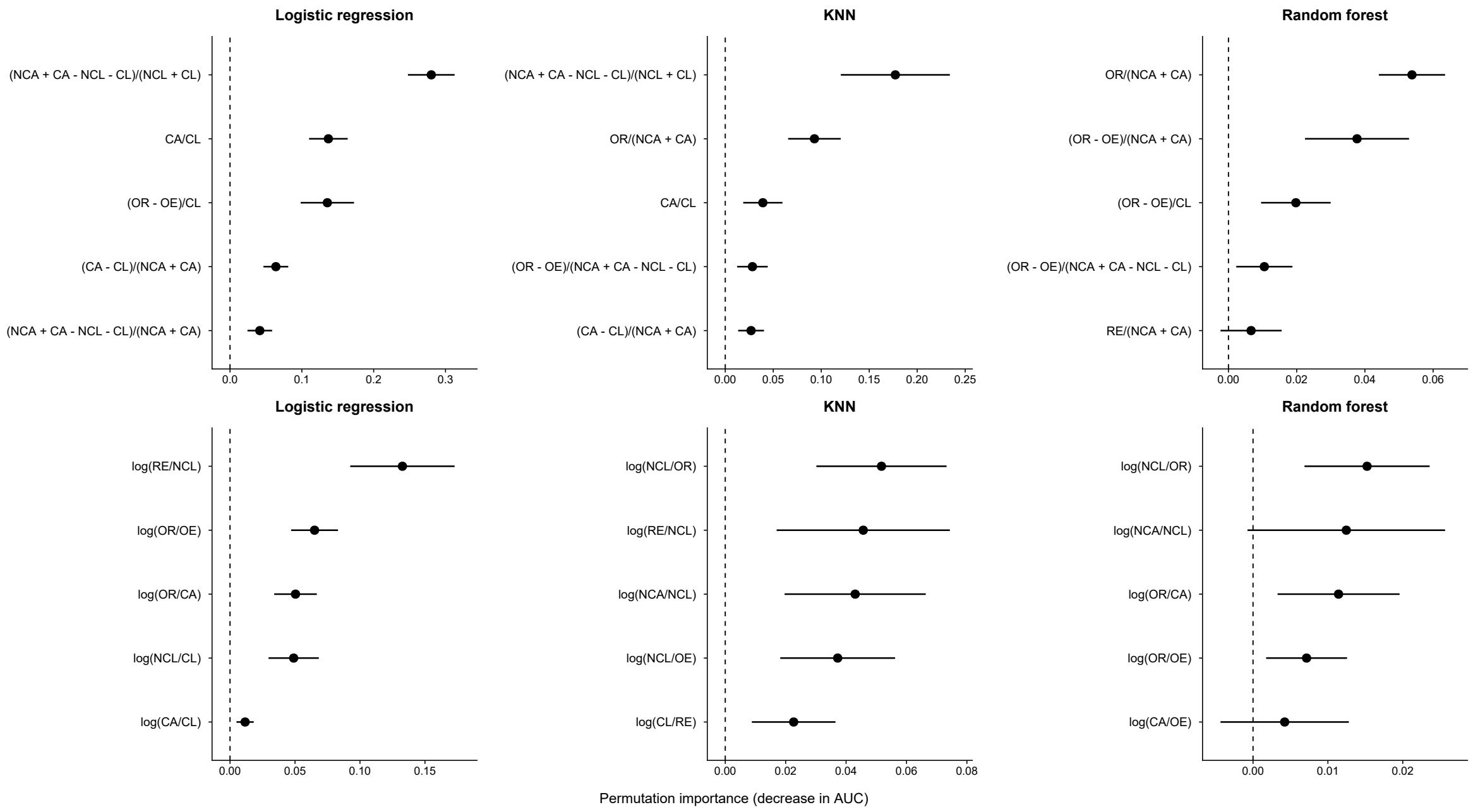

Figure 2: Five best variables according to permutation importance for logistic regression, k-nearest neighbors (KNN), and random forest models using standard ratios (top row) and compositional ratios (bottom row). Points indicate the mean decrease in the area under the receiver operating characteristic curve (AUC), and error bars represent ±1 standard deviation.

Table 2: Performance measures of the logistic regression, 5-nearest neighbours, and random forest models using standard and compositional ratio representations. Point estimates correspond to performance on the validation set, while values in parentheses indicate the 10th and 90th percentiles of the performance measures obtained across the repeated cross-validation resamples of the training set.

| Method | Accuracy | Sensitivity | Specificity | Balanced accuracy |
|---|---|---|---|---|
| LR (S) | 0.66 (0.61, 0.69) | 0.82 (0.57, 0.92) | 0.66 (0.61, 0.69) | 0.74 (0.62, 0.78) |
| LR (C) | 0.67 (0.63, 0.72) | 0.86 (0.64, 0.93) | 0.67 (0.63, 0.72) | 0.76 (0.65, 0.80) |
| KNN (S) | 0.70 (0.59, 0.67) | 0.75 (0.50, 0.86) | 0.70 (0.57, 0.68) | 0.73 (0.58, 0.74) |
| KNN (C) | 0.58 (0.57, 0.65) | 0.89 (0.57, 0.92) | 0.58 (0.56, 0.65) | 0.74 (0.59, 0.73) |
| RF (S) | 0.62 (0.59, 0.68) | 0.75 (0.57, 0.92) | 0.62 (0.61, 0.69) | 0.68 (0.62, 0.78) |
| RF (C) | 0.65 (0.64, 0.72) | 0.89 (0.64, 0.93) | 0.65 (0.64, 0.72) | 0.77 (0.66, 0.78) |

LR: Logistic regression; KNN: k-nearest neighbours; RF: Random; S: standard; C: Compositional

The area under the ROC curves in Figure 3 provides a similar picture, namely superiority of the compositional approach over standard ratios in logistic regression and random forests, and generally lower performance of k-nearest neighbours.

# 4. Discussion

Bankruptcy prediction is one of the most common uses of financial ratios and yet an under-researched area in the CoDa methodology. This article adapts Altman's prediction model and its most common variants to this methodology with three common prediction methods, one being a statistical model and the remaining two machine-learning methods, and compares the performance of standard and compositional financial ratios, computed as plr.

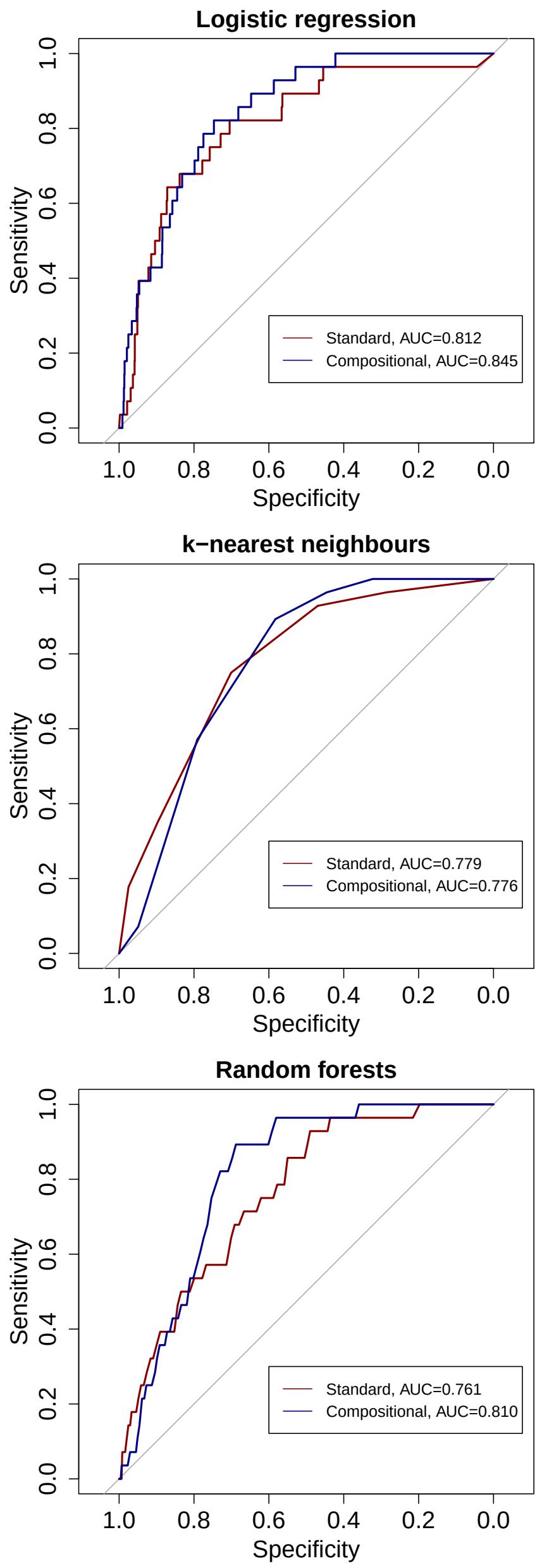

Figure 3: Receiver operating characteristic (ROC) curves and area under the curves (AUC)

As reported in the literature, the CoDa methodology improves on the skewness and kurtosis of the analysis variables. Hence, outlier pruning, which can improve the fulfilment of statistical assumptions but can also affect the sample representativity, is avoided. Unlike research done so far (Balzano and Magrini 2025; Magrini 2025; Pawełek and Sadko 2026; Martin-Melero *et al.* 2025), we have used the full sample for all methods.

The results show that beyond the fulfilment or not of statistical assumptions, the comparison between the standard and compositional approach is also a matter of predictive performance. Compositional methods are better in terms of sensitivity, with mixed results regarding specificity, compositional random forests and compositional logistic regression behaving the best. Magrini (2025) did not find any substantial difference between standard and compositional ratios for logistic regression and random forests, but this author pruned the extreme outliers for standard ratios. In spite of their known sensitivity to outliers, when it comes to predictive performance, logistic regression is just marginally worse for the outlier-rich standard ratios while k-nearest neighbours are actually better. In spite of their known robustness to outliers, when it comes to predictive performance, random forests behave much worse for standard ratios.

Implications for practitioners are twofold. From a methodological perspective, the article draws from the simple accounting figures used in Altman's model and shows how to easily compute the plr needed for the CoDa methodology, but any applied researcher can extend the idea to his or her favourite accounting figures, as long as they are non-negative. The conceptual simplicity of plr, compared to other log-ratio representations can encourage applied researchers to adopt the CoDa approach. From a financial perspective, in the context of the Spanish wholesale sector, the key ratios and accounting figures for bankruptcy prediction are turnover and non-current liabilities, which can be offset by retained earnings.

We are of course aware that the article's findings are not generalizable beyond this particular sector of the Spanish economy and the particular accounting figures and ratios chosen. This fact is compounded by the width of the intervals of the precision measures.

The possibility of using other log-ratio transformations than plr deserves a comment. Centred log-ratios a.k.a. clr (Aitchison 1983) are coherent with Aitchison's distances and therefore can be used for the k-nearest-neighbour method with identical results as the $D\,(D-1)\,/2$ plr. The family of isometric log-ratio transformations a.k.a. ilr (Egozcue, Pawlowsky-Glahn, Mateu-Figueras, and Barceló-Vidal 2003) has the same property: any ilr transformation can be used for the k-nearest-neighbour method with identical results as the clr and the $D\,(D-1)\,/2$ plr. Any ilr transformation can also be used as predictor in a logistic regression model with identical forecasts as the $D-1$ plr (Coenders and Pawlowsky-Glahn 2020). Tree-based methods in general and random forests in particular constitute another story. While the ilr family of transformations is a respectable choice (Magrini 2025), it must be borne in mind that the forecasts depend on the particular ilr selected within the family (Templ and Gonzalez-Rodriguez 2024), and that any of these forecasts also differ from those obtained with the $D\,(D-1)\,/2$ plr. The main advantage of plr is that there is only one way of selecting all possible $D\,(D-1)\,/2$ plr, which makes the results less arbitrary and always replicable (Engle and Chaput 2021; Tolosana-Delgado *et al.* 2019). On the contrary, using all possible ilr is generally unfeasible. With just $D = 7$ accounting figures there are 1,141 distinct balances for the ilr transformation and it is easy to imagine what can happen when $D$ is large. clr have also been used for random forests (Nathwani, Wilkinson, Fry, Armstrong, Smith, and Ihlenfeld 2022), being also replicable, but they yield different results from both ilr and plr. Tolosana-Delgado *et al.* (2019) reported superior performance of plr over ilr or clr for random forests. Since differences with respect to ilr will emerge only for random forests and Magrini (2025) used ilr with this algorithm, the Appendix shows the results of this method combination with our dataset.This provides direct evidence, in the specific context of bankruptcy prediction, of the relationship between the choice of log-ratio family and predictive accuracy for random-forests, the only one of the three methods for which such a relationship is not already guaranteed by theory.

Further research can include adapting other common machine-learning methods in bankruptcy prediction like *neural networks*, *deep learning*, *naïve Bayes classifier*, *support vector machines* (SVM), *natural gradient boosting* (NGBoost), *adaptative boosting* (AdaBoost), *categorical boosting* (CatBoost), *light gradient boosting* (lightGBM), *extreme gradient boosting* (XGBoost), and *metaheuristic algorithms.* The so-called *hybrid models* or *meta models* combine more than one machine-learning approach and are also worth exploring (Akusta and Gün 2025; Asif, Hemalatha, and Vivekananda Reddy 2025; Chohan, Li, Abrar, and Butt 2026; Hussain, Jingchun, Hussain, Shaique, Rehman, and Bhatti 2026; Idris and Ismail 2026; Veganzones *et al.* 2026). Each machine learning method can have a different set of compatible log-ratio transformations within the CoDa methodology, and this has to be explored in the first place. Besides, each machine learning method may be robust to outliers to a different extent, if at all.

The choice of the best method may be based not only on predictive quality but also on convenience and interpretability (Antar, Hassan, and Barka 2025; Chan, Tang, Tsai, and Yang 2025; Dam and Dang 2026; Dong and Li 2026; Gwalani *et al.* 2025; Hussain *et al.* 2026; Lin, Padliansyah, Lu, and Liu 2025; Molnar 2020, 2025; Rejeb, Rejeb, and Treiblmaier 2026).

Some methods provide an explicit and interpretable decision rule and some do not, some methods make it possible to select the best predictor subset and some do not, some are well known to a wide practitioner audience and some are not, or not to the same extent. When machine learning methods are not interpretable ante-hoc, interpretability of machine learning results can be enhanced post-hoc by means of the so-called *explainable artificial intelligence* (XAI) methods, *Shapley additive explanations* –SHAP– (Antar *et al.* 2025; Chan *et al.* 2025; Chao, Elias, Yahya, and Jenal 2025; Dong and Li 2026; Hussain *et al.* 2026; Kim and Sun 2026; Lin *et al.* 2025; Malhotra, Malhotra, Nydick, and Coates 2026; Yang, Zhang, Li, Hong, and Wang 2026; Zhou 2026) and *local interpretable model-agnostic explanations* –LIME– (Antar *et al.* 2025; Lin *et al.* 2025; Zhang, Chen, and Tian 2026) being the most popular.

Additional avenues for further research are the inclusion of distributed lags of the explanatory variables and the exploration of different forecasting horizons (Magrini 2025), and the use of additional accounting figures beyond the $D = 7$ used in Altman's model and its derivatives (Nugroho and Dewayanto 2025).

A strong limitation of our study is that our sample had a very small number of bankrupt firms. Using a larger portion of the country's economy than the 2-digit NACE classification, or even the whole economy (Martin-Melero *et al.* 2025), is a possibility, although predictions may be industry-specific (Balzano and Magrini 2025). Taking together all trade sectors may be a wise compromise. This would narrow the intervals of the performance measures, which overlap in Table 2.

## Acknowledgments

This work was supported by the Spanish Ministry of Science, Innovation and Universities MCIN/AEI/10.13039/501100011033 and ERDF-a way of making Europe (grant number PID2021-123833OB-I00); the Spanish Ministry of Health (grant number CIBERCB06/02/1002); and AGAUR and the Department of Climate Action, Food and Rural Agenda of Generalitat de Catalunya (grant number 2023-CLIMA-00037).

## Conflicts of interest

The authors declare no conflict of interest.

## Data availability

The curated data are available from the corresponding author upon reasonable request.

# A. Results of random forests with ilr

In this appendix we compare the predictive performance of ilr and plr for the random-forest algorithm. Of the many possible ilr transformation we choose the following one, which is interpretable in financial terms and contains three log-ratios which are equivalent to those in Equation (3) after scaling:

$$\begin{aligned}
&\text{turnover}: && \sqrt{\tfrac{2\times5}{2+5}}\log\left(\frac{(OR\times OE)^{1/2}}{(NCA\times CA\times RE\times NCL\times CL)^{1/5}}\right),\\
&\text{margin}: && \sqrt{\tfrac{1\times1}{1+1}}\log\left(\frac{OR}{OE}\right),\\
&\text{retained earnings over balance sheet}: && \sqrt{\tfrac{1\times4}{1+4}}\log\left(\frac{RE}{(NCA\times CA\times NCL\times CL)^{1/4}}\right),\\
&\text{solvency}: && \sqrt{\tfrac{2\times2}{2+2}}\log\left(\frac{(NCA\times CA)^{1/2}}{(NCL\times CL)^{1/2}}\right),\\
&\text{asset tangibility}: && \sqrt{\tfrac{1\times1}{1+1}}\log\left(\frac{NCA}{CA}\right),\\
&\text{debt maturity}: && \sqrt{\tfrac{1\times1}{1+1}}\log\left(\frac{NCL}{CL}\right).
\end{aligned} \tag{5}$$

The area under the ROC curve for the ilr was nearly identical to that for the plr at 0.805. The remaining precision measures and their corresponding intervals are in Table 3. The last row of Table 2 referring to the plr has been added atop the ilr results to ease comparison. These results are considerably worse than those with plr for sensitivity, and slightly worse for the remaining metrics.

Table 3: Performance measures of the random forest models using plr and ilr representations. Point estimates correspond to performance on the validation set, while values in parentheses indicate the 10th and 90th percentiles of the performance measures obtained across the repeated cross-validation resamples of the training set.

| Method | Accuracy | Sensitivity | Specificity | Balanced accuracy |
|---|---|---|---|---|
| RF (plr) | 0.65 (0.64, 0.72) | 0.89 (0.64, 0.93) | 0.65 (0.64, 0.72) | 0.77 (0.66, 0.78) |
| RF (ilr) | 0.64 (0.62, 0.69) | 0.79 (0.54, 0.93) | 0.64 (0.61, 0.69) | 0.71 (0.61, 0.78) |

It must be taken into account that changing the ilr transformation from that in Equation (5) would change the predictive performance metrics. Conversely, any ilr would lead to the same predictions as plr for logistic regression and k-nearest neighbours.

## Affiliation:

Germà Coenders
Faculty of Economics and Business
C/ Universitat de Girona 10. 17003 Girona, Spain
E-mail: germa.coenders@udg.edu
URL: https://www.udg.edu/ca/directori/pagina-personal?ID=2000949